# Acoustic and Optical Phonons in Quasi-Two-Dimensional MPX$_3$ Van der Waals Semiconductors

Dylan Wright,[1,2] Zahra Ebrahim Nataj,[1,2] Erick Guzman,[1] Jake Polster,[3] Menno Bouman,[3] Fariborz Kargar,[1,2,4,*] Alexander A. Balandin[1,2,*]

[1]Department of Materials Science and Engineering, University of California, Los Angeles, California 90095 USA

[2]California NanoSystems Institute, University of California, Los Angeles, California 90095 USA

[3]KLA Instruments Group, West Covina, California 91791 USA

[4]Materials Research and Education Center, Department of Mechanical Engineering, Auburn University, Auburn, Alabama 36849 USA

We report the results of the investigation of the acoustic and optical phonons in quasi-two-dimensional antiferromagnetic semiconductors of the transition metal phosphorus trisulfide family with Mn, Fe, Co, Ni, and Cd as metal atoms. The Brillouin-Mandelstam and Raman light scattering spectroscopies were conducted at room temperature to measure the acoustic and optical phonon frequencies close to the Brillouin zone center along the $\Gamma - A$ high symmetry direction. The absorption and index of refraction were measured in the visible and infrared ranges using the reflectometry technique. We found an intriguing large variation, over ~28%, in the acoustic phonon group velocities in this group of materials with similar crystal structures. Our data indicate that the full-width-at-half-maximum of the acoustic phonon peaks is strongly affected by the optical properties and the electronic band gap. The acoustic phonon lifetime extracted for some of the materials was correlated with their thermal properties. The obtained results are important for understanding the layered van der Waals semiconductors and for assessing their potential for optoelectronic and spintronic device applications.

---

[*] Corresponding authors: f.kargar@ucla.edu (F.K.); balandin@seas.ucla.edu (A.A.B.)

Recently, low-dimensional magnetic semiconductors have become a focus of theoretical and experimental investigation as a new group of materials with potential applications in spintronic devices.[1–9] Following the discovery of graphene, a two-dimensional (2D) single-atomic sheet of carbon atoms, interest was directed to magnetic graphene-like structures that host different forms of spin order even at few and single-layer limits.[2,3,10] Metal-transition phospho-trichalcogenides, referred to as $MPX_3$ in which M is a transition metal and X is a chalcogen, is a class of compounds that attracted much attention. Thin films of many of such materials with the few-atomic-plane and the single-atomic-plane thicknesses reveal antiferromagnetic (AFM) spin configurations, below their respective Néel temperatures, $T_N$.[11–16] Theoretical predictions suggest that quasi-2D $MPX_3$ materials have spin order tunability *via* external stimuli such as strain fields and electric gating.[11,17] The latter combined with their low cleavage energy present opportunities for the fabrication of magnetic heterostructures with the potential for electronically controlled spintronic technologies.[18,19] While the properties of AFM electrical conductors and insulators are well known, the AFM semiconductors are still poorly understood. Relatively little is known about the interplay of the electron, magnon, and phonon properties of such materials.

Early interest in $MPX_3$ compounds was motivated by their prospects for applications as cathodes in lithium batteries with research focused on characterization of the material properties in their bulk form.[20] There are published reports on bulk magnetic, electrical, and optical phonon properties of $MPX_3$.[21–23] However, the information on the acoustic phonon properties of many $MPX_3$ materials is missing.[24] Acoustic phonons carry heat and limit the electron mobility near room temperature in semiconductor materials. The lattice thermal conductivity, $K_P$, of semiconductors is directly related to the group velocity of acoustic phonons, $v$, via $K_p \sim Cv\Lambda$, where $C$ and $\Lambda$ represent the volumetric specific heat and phonon mean free path, respectively. The characteristics of the long-wavelength acoustic phonons are defined by the elastic coefficients of the media.[25] The latter is particularly important considering that the magnetic states of $MPX_3$ films can be engineered through strain fields.[11,19,26] The practical value of the knowledge of acoustic phonons is not limited to heat transport, scattering of electrons, and determining mechanical properties but also extends to the understanding of magnon–phonon scattering. Theoretical studies have shown that the interplay between low-energy magnons and long-wavelength acoustic phonons close to the Brillouin zone (BZ) center influences the magnon relaxation processes.[27]

In this study, we report the results of the investigation of the acoustic and optical phonons in a set of $MPX_3$ compounds with Mn, Fe, Co, Ni, and Cd transition metal atoms. We aim to determine the phonon characteristics of the AFM semiconductors important for understanding their properties. The questions we ask are: How do the details of the crystal structure affect the acoustic and optical phonon spectra in this type of material? Can we observe any signatures of the interplay of the electronic, phononic, and spin properties of these materials using optical spectroscopy tools? To achieve this, we employ micro-Brillouin-Mandelstam light scattering (BMS) and micro-Raman techniques to detect acoustic and optical phonons, respectively. BMS is a nondestructive optical technique that is used to detect elemental excitations such as acoustic phonons and magnons with energies in the range from 2 GHz to 900 GHz and wavevectors close to the BZ center.[28] The optical properties of $MPX_3$ films were measured *via* spectral reflectometry in the visible light range. The data on the optical properties are required for the interpretation of the BMS spectra and for extracting the group velocity of acoustic phonons. The high-quality bulk single crystals of $MnPS_3$, $FePS_3$, $CoPS_3$, $NiPS_3$, and $CdPS_3$ crystals were synthesized using the chemical vapor transition (CVT) method (2D Semiconductors, USA). Raman and BMS measurements were performed on bulk crystals at room temperature (RT). The crystals were mechanically exfoliated on $Si/SiO_2$ substrates for optical reflectometry measurements. The thickness of the exfoliated samples ranged about a few hundred nanometers while their lateral dimensions typically varied between 10 μm to 30 μm.

Figure 1 presents a schematic of the different magnetic order and crystal structures of the studied materials. The spin arrangement in these semiconductors is primarily influenced by the size of the "M" elements which are arranged in hexagonal sublattices. The diverse ionic radii of different transition metal cations lead to various AFM spin configurations such as Néel, zigzag Ising, or zigzag XY spin ordering.[29] Figure 1 (a) illustrates different magnetic arrangements observed in different compounds within this family. The small ionic radii for the $Co^{2+}$ and $Ni^{2+}$ cations result in flat octahedral structures around the transition metal, where spins are constrained from freely rotating about the *ab* plane.[29] In contrast, the larger radii of $Fe^{2+}$ and $Mn^{2+}$ causes the spins to rotate about the *ab* plane and align along the *c*-axis of these compounds. As a result, $MnPS_3$ and $FePS_3$ exhibit out-of-plane spins with Néel and zigzag Ising AFM alignment, respectively, while $CoPS_3$ and NiPS$_3$ both have zigzag XY ordering with spins oriented within the *ab* plane.[29] Other compounds with the metal elements of Cd and Zn are nonmagnetic. The effect of ionic radius can be seen in the Néel temperature as well. Compounds with the smallest cation radius and lattice constants show the highest Néel temperatures.[11,29] All MPX$_3$s have a monoclinic structure, comprised of three unequal axis lengths with two being perpendicular and the third at an oblique angle to the perpendicular axes. They all have two-fold, body-centered mirror symmetry corresponding to the C/2m space group. The sulfur atoms bond with phosphorus to form dumbbell-like units of P$_2$S$_6$. Figure 1 (b-f) presents the crystal structure and spin alignment of different MPX$_3$ compounds. Table 1 summarizes the lattice parameters, Néel temperatures, and AFM type of each compound used in this study.

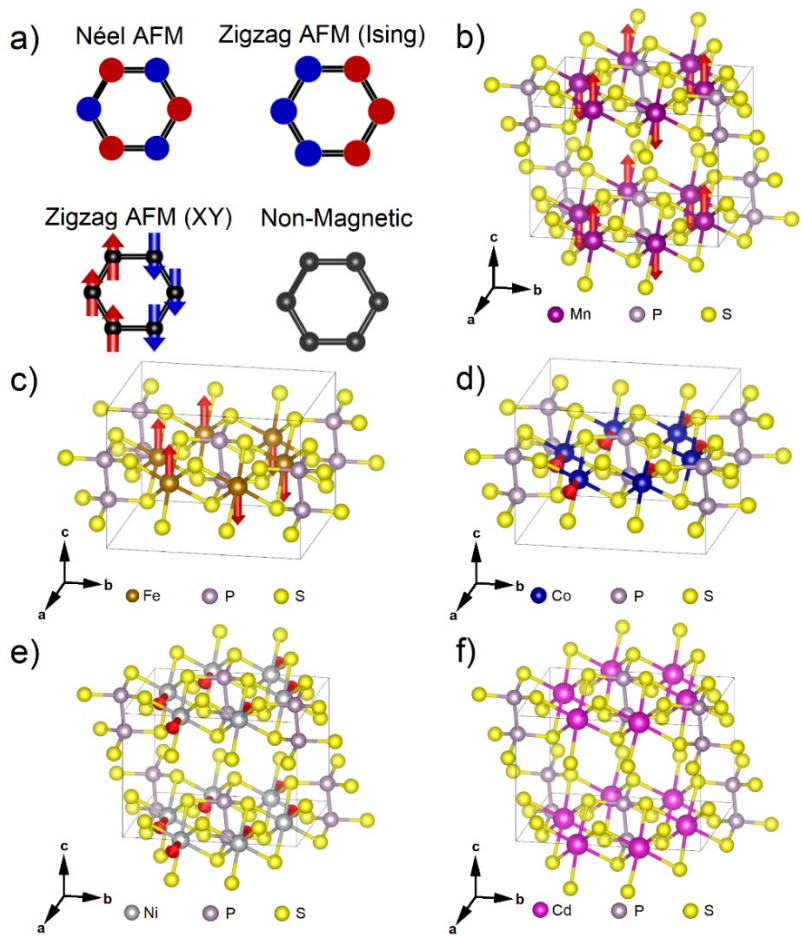

[**Figure 1**: (a) Schematic showing different magnetic configurations of MPX$_3$ semiconducting compounds. The blue and red spheres represent opposite spin ordering perpendicular to the *ab* plane. The arrows indicate the opposite spin ordering within the *ab* plane. The crystal structures of various MPX$_3$s with metal atoms of (b) Mn, (c) Fe, (d) Co, (e) Ni, and (f) Cd. The arrows show the spin orientation configuration in different crystals.]

**Table 1**: Structural, electronic, and magnetic properties of MPX$_3$ compounds

|  | a (Å) | b (Å) | c (Å) | β (°) | Bandgap (eV)* | Spin Order | T$_N$ (K) | Refs. |
|---|---|---|---|---|---|---|---|---|
| MnPS$_3$ | 6.077 | 10.524 | 6.796 | 107.35 | 3.0 (D) | Neel AFM | 78 | 11,12,30,31 |
| FePS$_3$ | 5.947 | 10.300 | 6.7222 | 107.16 | 1.5 (I) | Zigzag (Ising) | 118 | 10–12 |
| CoPS$_3$ | 5.901 | 10.222 | 6.658 | 107.17 | 1.6 (I) | Zigzag (XY) | 132 | 11,12,32 |
| NiPS$_3$ | 5.812 | 10.070 | 6.632 | 106.98 | 1.4 (I) | Zigzag (XY) | 155 | 11,12,33 |
| CdPS$_3$ | 6.218 | 10.763 | 6.867 | 107.58 | 3.3 (I) | NM | NM | 11,12 |

*"D" and "I" stands for direct and indirect bandgap materials.

Raman spectroscopy was performed in the conventional backscattering configuration using two laser excitation wavelengths of 488 nm (blue) and 633 nm (red) at RT. The Raman filter cutoff wavenumber for both blue and red lasers was ~110 cm$^{-1}$. In all experiments, the incident light was perpendicular to the (001) surface. The polarization of the scattered light was not analyzed. To avoid any local heating effects, the excitation laser power was kept below ~500 µW in all measurements. Figure 2 (a, b) shows the Raman spectra of all samples under 488 nm and 633 nm laser excitations, respectively. Note that at RT, all MPX$_3$s are in the paramagnetic phase and therefore, they belong to the $C_{2h}$ point symmetry group with the irreducible representation of $\Gamma = 8A_g + 6A_u + 7B_g + 9B_u$. Among these modes, $A_g$ and $B_g$ modes are Raman-active which are assigned to their respective Raman peaks in Figure 2 (a, b).[22] Note that the symmetry of these compounds would be different in their monolayer form below their magnetic transition with all nonmagnetic and Néel antiferromagnetic types belonging to $D_{3d}$ and the zigzag AFMs preserving their $C_{2h}$ point symmetry groups.[12]

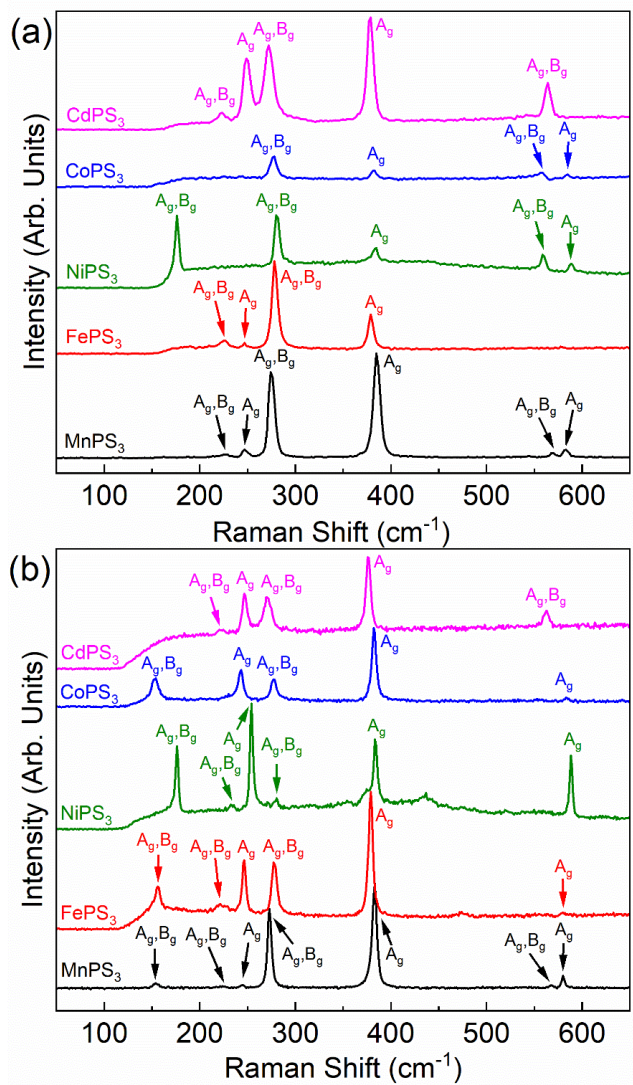

[Figure 2: Raman spectra of MPX$_3$ layered semiconductor compounds at room temperature using different laser excitations of (a) 488 nm, and (b) 633 nm with the vibrational mode symmetries assigned.]

It is important to mention that not all $A_g$ and $B_g$ modes were detectable in the spectra measured for all compounds, either due to their weak intensities or Raman filter cutoff limitations in the low-wavenumber range. For example, we were not able to detect the expected $A_g$ mode at ~580 cm$^{-1}$ for CdPS$_3$. The spectral frequency and the full-width-at-half-maximum (FWHM) of the Raman peaks along with their mode assignments are listed in Table 2. It is generally considered that despite the complex atomic structure of MPX$_3$ materials, their vibrational characteristics can be pictured into isolated high-wavenumber vibrations of P$_2$S$_6^{4-}$ and lower-wavenumber vibrations of M$^{2+}$ and P$_2$S$_6^{4-}$ with spectral features above and below ~200 cm$^{-1}$, respectively.[22,24] As a result, the spectral position of the peaks in the high wavenumber ranges remains largely invariant across structures with different M atoms.[22] In contrast, the spectral position of the peaks at low wavenumber ranges is influenced by the transition metal. While the degenerate $A_g, B_g$ modes for MnPS$_3$, FePS$_3$, and CoPS$_3$ are observed at ~155 cm$^{-1}$, it is located at 176 cm$^{-1}$ for NiPS$_3$. The reason for such a large spectral shift given the almost identical atomic masses of Mn, Fe, Ni, and Co is most likely related to a smaller cation radius of Ni$^{+2}$ and requires theoretical calculations. The atomic mass of Cd is almost twice that of the other transition metals of interest in this study and thus, the same degenerate peaks are expected to be situated at ~110 cm$^{-1}$. Prior Raman studies of CdPS$_3$ confirm a degenerate $A_g, B_g$ modes at or close to the calculated wavenumber.[34] It is worth noting that certain reports attribute the Raman peaks of MPX$_3$ structures in accordance to the $D_{3d}$ point group symmetry due to the weak interlayer weak vdW bonds, with $E_g$ and $A_{1g}$ modes being Raman active.[34] This assumption has led to inconsistencies in Raman mode assignment across the literature. However, we contend that, considering the $C_{2h}$ point symmetry of these structures, Raman-active modes should be designated as $A_g$ and $B_g$.

**Table 2:** Frequency and FWHM of Raman peaks along with their vibrational symmetries

| | Symmetry | | $A_g, B_g$ | $A_g, B_g$ | $A_g$ | $A_g, B_g$ | $A_g$ | $A_g, B_g$ | $A_g$ |
|---|---|---|---|---|---|---|---|---|---|
| MnPS$_3$ | Frequency | Red | 154.1 | 223.7 | 244.6 | 273 | 382.8 | 567.6 | 580.2 |
| | | Blue | - | 226.4 | 247.5 | 275.2 | 385.2 | 569.3 | 583 |
| | FWHM | Red | 6.5 | 5.6 | 4.3 | 5.7 | 6.6 | 6.3 | 4.3 |
| | | Blue | - | 7.5 | 6.7 | 8 | 9.1 | 7.5 | 8.1 |
| FePS$_3$ | Frequency | Red | 155.8 | 223.2 | 246.4 | 278 | 378.8 | - | 580.2 |
| | | Blue | - | 225.2 | 246.7 | 278.5 | 379.0 | - | - |
| | FWHM | Red | 7.4 | 11.4 | 4.9 | 6.9 | 6.3 | - | 5.8 |
| | | Blue | - | 9.7 | 3.2 | 8.0 | 7.3 | - | - |
| NiPS$_3$ | Frequency | Red | 176.1 | 233.2 | 254.2 | 279.8 | 383.7 | - | 588.4 |
| | | Blue | 176.1 | - | - | 280.9 | 383.3 | 559.4 | 588.8 |
| | FWHM | Red | 3.8 | 4.7 | 4.0 | 4.0 | 6.3 | - | 3.8 |
| | | Blue | 5.4 | - | - | 6.6 | 7.8 | 6.1 | 5.2 |
| CoPS$_3$ | Frequency | Red | 153.1 | - | 242.6 | 277.4 | 382.3 | - | 583.9 |
| | | Blue | - | - | - | 277.2 | 382.1 | 556.0 | 584.5 |
| | FWHM | Red | 9.9 | - | 8.4 | 8.1 | 6.4 | - | 4.6 |
| | | Blue | - | - | - | 9.4 | 7.3 | 10.2 | 5.1 |
| CdPS$_3$ | Frequency | Red | - | 221.9 | 247.1 | 271 | 376.3 | 562.5 | - |
| | | Blue | - | 222.9 | 249.4 | 272.2 | 378.4 | 564 | - |
| | FWHM | Red | - | 8.6 | 6.2 | 9.4 | 6.5 | 6.8 | - |
| | | Blue | - | 10.7 | 10.0 | 12.9 | 8.4 | 7.8 | - |

The optical properties, *i.e.*, the index of refraction, $n$, and extinction coefficient, $k$, of the samples were determined using spectral reflectometry (Filmetrics F40, KLA Instruments, USA). In these experiments, the samples were illuminated with a tungsten-halogen light source, and the reflectance was measured as a function of the wavelength in the range of 400 to 1000 nm. The measurements were performed with the incident light normal to the sample surface. The spot size on the sample was ~8 µm. Fittings were performed to determine the refractive index and optical extinction coefficient as a function of wavelength. The results for $n$ and $k$ are presented in Figure 3 (a, b). The examined materials show a general trend of decreasing refractive index with the wavelength. Semiconductors with smaller band gaps exhibit higher refractive indices, as expected.[35] It is known that $MPX_3$s are semiconductors with bandgaps spanning the mid-range of 1.3 eV to a wide-bandgap range of 3.5 eV (see Table 1).[35] All structures in this study are indirect bandgap semiconductors except for $MnPS_3$ which is a direct bandgap semiconductor. The data on the optical properties at 532 nm excitation wavelength is required for the interpretation of BMS spectra which will be discussed next.

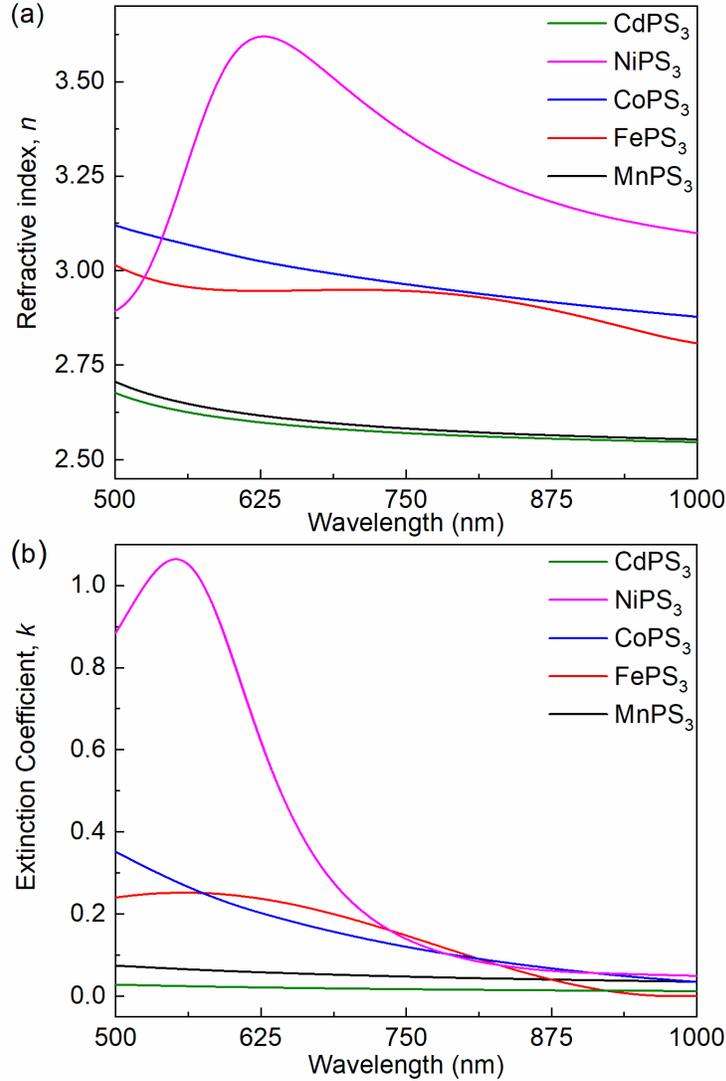

[Figure 3: (a) Index of refraction, $n$, and (b) extinction coefficient, $k$ of MPX$_3$ compounds at room temperature.]

We now turn to the main element of this work which is the investigation of the acoustic phonon energies in MPX$_3$ compounds using the BMS technique. The BMS measurements were conducted in the backscattering configuration using a continuous wave diode-pumped laser (Spectra-Physics, Excelsior, USA) with an excitation wavelength of $\lambda = 532$ nm. The incident laser power was measured at ~10 mW at a constant incident angle of $\theta = 45°$ on all samples. The incident and scattered light were both $p$-polarized. The details of the BMS setup and experimental procedure can be found in our prior reports for other material systems.[36–38] The results are presented in Figure 4. The spectra are vertically transferred for better visualization. Two sharp peaks on both sides of the spectra, corresponding to the Stokes and anti-Stokes processes, were observed in the frequency range of ±27 GHz to ±35 GHz, for all samples. These peaks depicted by "LA" in the plot are attributed to longitudinal acoustic phonons propagating close to the $c$ crystallographic axis in the real space, or the $\Gamma - A$ direction in the reciprocal space, for all structures. Symmetrical spurious peaks were also observed in the Stokes and anti-Stokes sides of the spectra that are shaded by grey rectangles for clarification. These peaks likely originated from the optical cavity formed by the reflection of the light from the front and back surfaces of the samples.

The spectral frequency, $f$, and the FWHM of the LA peaks were determined by fitting individual Lorentzian functions to the experimental data. The obtained frequencies for MPX$_3$s with Mn, Fe, Co, Ni, and Cd transition metal atoms are 33 GHz, 34 GHz, 37 GHz, 33 GHz, and 27 GHz, respectively. Note that, unlike optical phonons which typically have a flat dispersion in the vicinity of the Brillouin zone center, acoustic phonons have a linear dispersion as $\omega = qv$ close to the $\Gamma$ point, in which $q = 4\pi n/\lambda$ is the phonon wavevector, $\omega = 2\pi f$ is the angular frequency, and $v$ is the phonon group velocity, respectively. Using the dispersion relation, the group velocity can be calculated according to $v = f\lambda/2n$. We obtained a group velocity of 2960 m/s, 2713 m/s, 3203 m/s, 2310 m/s, and 3077 m/s for the Mn, Fe, Co, Ni, and Cd-based compounds, respectively. The results show that longitudinal acoustic phonons propagate with rather slow velocities in the range of ~2300 to ~3200 m/s for MPX$_3$ compounds, with NiPS$_3$ and CoPS$_3$ exhibiting the slowest and fastest propagation velocities, respectively. The obtained results are in acceptable agreement with limited studies investigating the acoustic phonon properties in layered materials.[39] One prior Brillouin spectroscopy study of the acoustic phonons in CdPS$_3$ reported a peak associated with LA phonons propagating close to the $c$ crystallographic direction at 28 GHz which is consistent with our results.[39] However, the reported $v$ in that study was slightly lower than our value owing to the higher $n$, which was taken from the literature.

An intriguing observation in the spectra presented in Figure 4 is a substantial disparity among the FWHM of the LA peaks for different compounds. The large FWHM of the BMS peaks in specific compounds is a direct manifestation of high optical absorption in those structures.[40] In fact, as seen in Figure 4, the LA peaks in the spectra accumulated for Fe, Ni, and Co-based structures are so broad that they are barely detectable even after long accumulation times. For an opaque or a semitransparent crystal with the complex refractive index $\eta = n + ik$, an optical broadening of $\delta\omega/\omega = 2k/n$ occurs due to the scattering of light from phonons with momentums within $\delta p \sim 2kk_0\hbar$ and mean phonon momentum of $p = q\hbar$ due to the uncertainty principle.[40] For Fe, Ni, and Co-based structures, taking the measured values of $n$ and $k$, the optical broadening is calculated as ~5.7 GHz, 22.5 GHz, and 7.2 GHz, respectively. However, the optical broadening for semi-transparent materials of MnPS$_3$ and CdPS$_3$ at $\lambda = 532$ nm is ~1.7 GHz and 0.6 GHz. The experimental FWHM of the LA peaks obtained from individual Lorentzian fittings on the BMS data yields ~1.0 GHz, ~6.5 GHz, ~18 GHz, 1.6 GHz, and 1.7 GHz for Mn, Fe, Co, Ni, and Cd-based structures, respectively. Comparing the experimental and calculated optical broadening values, one would infer that only in the case of CdPS$_3$ the optical broadening is significantly lower than the obtained experimental FWHM. In this specific case, the FWHM can be related to the phonon lifetime considering the energy-time uncertainty relation $\Delta E/\hbar = 1/\tau$ in which $\Delta E$, $\hbar$, and $\tau$ are the BMS peak FWHM [rad/s], reduced Planck's constant, and the phonon lifetime, respectively. A phonon lifetime of ~144 ps is derived after subtracting the optical broadening of 0.6 GHz from the measured 1.7 GHz FWHM for the probed LA phonon in CdPS$_3$. The obtained phonon lifetime is rather short for such long-wavelength acoustic phonons with wavevectors close to the BZ center and frequency of 28 GHz. To provide a comparison, the reported lifetime of 50 to 100 GHz LA phonons in a silicon wafer is ~5-7 ns while for 30 GHz LA phonons in free-standing silicon membranes is ~2 ns.[41,42] The slow phonon group velocity and short phonon lifetime of these structures imply a poor cross-plane thermal conductivity of these compounds. A prior experimental study reports significantly low cross-plane thermal conductivity values of 0.85 W/mK and 1.1 W/mK for FePS$_3$ and MnPS$_3$, respectively, which aligns with our findings on acoustic phonon properties.[24]

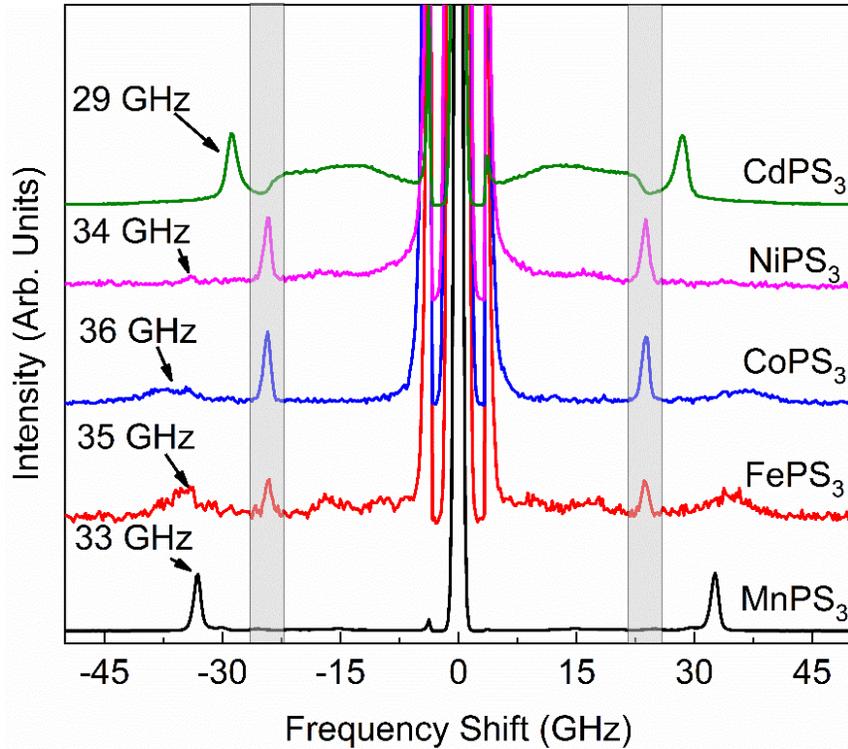

[Figure 4: BMS spectra of MPX$_3$ compounds performed at room temperature at an incidence angle of 45°. Peaks shown with arrows are associated with the longitudinal acoustic (LA) phonon modes.]

At cryogenic temperatures, long-wavelength acoustic phonons are the main heat carriers in semiconductor materials. Lattice thermal conductivity, $K_p \sim Cv\Lambda$, is directly related to the phonon group velocity. An intuitive method to verify the accuracy of group velocities obtained through BMS measurements is to examine the thermal conductivity of these compounds at cryogenic temperatures. Such data is available for FePS$_3$, MnPS$_3$, and NiPS$_3$.[43] Our findings show that the group velocity of NiPS$_3$ is lower than that of FePS$_3$, which in turn is slightly less than that of MnPS$_3$. Interestingly, existing thermal transport data also supports this trend, showing that NiPS$_3$ and MnPS$_3$ have the smallest and largest thermal conductivity values, respectively. Our results on the acoustic phonons, *i.e.,* phonon group velocities and the phonon lifetime, correlated with the available thermal data, suggest that the AFM semiconductors would require special thermal management provisions for practical applications. The latter may include placing MPX$_3$ on thermally conductive substrates, *e.g.*, diamond, in a similar way as it is done for ultra-wide-band-gap Ga$_2$O$_3$ semiconductors characterized by low thermal conductivity.

In conclusion, we reported the results of the investigation of the acoustic and optical phonons in quasi-2D layered semiconductors of the MPX$_3$ family with Mn, Fe, Co, Ni, and Cd as metal atoms. All examined materials, except one CdPS$_3$, are AFM materials when cooled down below their respective Néel temperatures. We used BMS and Raman spectroscopy to measure the acoustic and optical phonon frequencies at RT. We found a large variation, over ~28%, in the acoustic phonon group velocities in this group of materials with similar crystal structures. Our data indicate that the width of the acoustic phonon peaks is strongly affected by the optical properties and electronic band gap. The acoustic phonon lifetime extracted for some of the materials was correlated with their thermal properties. The obtained results are important for understanding the layered antiferromagnetic van der Waals semiconductors and for assessing their potential for optoelectronic and spintronic device applications.

**Acknowledgment**

This work was supported, in part, by the National Science Foundation (NSF), Division of Material Research (DMR) *via* the project No. 2205973 entitled "Controlling Electron, Magnon, and Phonon States in Quasi-2D Antiferromagnetic Semiconductors for Enabling Novel Device Functionalities." A.A.B. and F.K. acknowledge the support of the National Science Foundation (NSF) *via* a Major Research Instrument (MRI) DMR project No. 2019056 entitled "Development of a Cryogenic Integrated Micro-Raman-Brillouin-Mandelstam Spectrometer." The authors thank Dr. Subhajit Ghosh, UCLA for useful discussions regarding the optical properties of $MPX_3$ structures.

**Conflict of Interest**

The authors declare no conflict of interest.

**Author Contributions**

A.A.B. and F.K. coordinated the project and led the data analysis and manuscript preparation; D.W. conducted Brillouin spectroscopy and contributed to data analysis; Z.E.N. conducted Raman spectroscopy; J.P. and M.B. performed refractive index and absorption measurements.; E.G. assisted with Brillouin measurements. All authors contributed to the manuscript preparation.

**The Data Availability Statement**

The data that support the findings of this study are available from the corresponding author upon reasonable request.

14 | P a g e

Lett. **28**(4), 237–240 (1972).

[41] B.C. Daly, K. Kang, Y. Wang, and D.G. Cahill, "Picosecond ultrasonic measurements of attenuation of longitudinal acoustic phonons in silicon," Phys. Rev. B - Condens. Matter Mater. Phys. **80**(17), 174112 (2009).

[42] J. Cuffe, O. Ristow, E. Chávez, A. Shchepetov, P.O. Chapuis, F. Alzina, M. Hettich, M. Prunnila, J. Ahopelto, T. Dekorsy, and C.M. Sotomayor Torres, "Lifetimes of confined acoustic phonons in ultrathin silicon membranes," Phys. Rev. Lett. **110**(9), 095503 (2013).

[43] A. Haglund, Thermal conductivity of MXY$_3$ magnetic layered trichalcogenides, The University of Tennessee, Knoxville, 2019.19 | P a g e